\documentclass[journal]{IEEEtran}
 \usepackage{amsmath,amssymb}
 \usepackage{subfigure}
 \usepackage{graphicx,graphics,color,psfrag}
 \usepackage{cite,balance}
 \usepackage{algorithm}
 \usepackage{accents}
 \usepackage{amsthm}
 \usepackage{bm}
 \usepackage{url}
 \usepackage{algorithmic}
 \usepackage[english]{babel}
 \usepackage{multirow}
 \usepackage{enumerate}
 \usepackage{cases}
 \usepackage{stfloats}
 \usepackage{dsfont}
 \usepackage{color,soul}
 \usepackage{amsfonts}
  \usepackage{tcolorbox}
 \usepackage{cite,graphicx,amsmath,amssymb}
 \usepackage{subfigure}
 \usepackage{fancyhdr}
 \usepackage{hhline}
 \usepackage{graphicx,graphics}
 \usepackage{array,color}
 \usepackage{amsmath}
 \usepackage{amsthm}

\include{header}

\IEEEoverridecommandlockouts

\begin{document}

\title{Edge Perception: Intelligent Wireless Sensing at Network Edge }
\author{Yuanhao Cui, Xiaowen Cao, Guangxu Zhu, Jiali Nie, and Jie Xu
\thanks{Y. Cui is with Southern University of Science and Technology, Shenzhen, China (e-mail: yuanhao.cui@ieee.org).}
\thanks{X. Cao is with the College of Electronics and Information Engineering, Shenzhen University, Shenzhen 518060, China (e-mail: caoxwen@szu.edu.cn). X. Cao is the corresponding author.}
\thanks{G. Zhu is with Shenzhen Research Institute of Big Data, Shenzhen 518172, China (e-mail: gxzhu@sribd.cn). }
\thanks{J. Nie is with  Beijing University of Posts and Telecommunications, Beijing, China (email:niejl@bupt.edu.cn).}
\thanks{J. Xu is with the School of Science and Engineering and the Future Network of Intelligence Institute, The Chinese University of Hong Kong (Shenzhen), Shenzhen 518172, China (e-mail: xujie@cuhk.edu.cn). }
}

\markboth{}{}
\maketitle

\setlength\abovedisplayskip{2pt}
\setlength\belowdisplayskip{2pt}

\vspace{-1.5cm}

\begin{abstract}
Future \emph{sixth-generation} (6G) networks are envisioned to support intelligent applications across various vertical scenarios, which have stringent requirements on high-precision sensing as well as ultra-low-latency data processing and decision making. Towards this end, a new paradigm of edge perception networks emerges, which integrates wireless sensing, communication, computation, and \emph{artificial intelligence} (AI) capabilities at network edge for intelligent sensing and data processing.  This article provides a timely overview on this emerging topic. We commence by discussing wireless edge perception, including physical layer transceiver design, network-wise cooperation, and  application-specific data analytics, for which the prospects and challenges are emphasized.
Next, we discuss the interplay between edge AI and wireless sensing in edge perception, and present various key techniques for two paradigms, namely edge AI empowered sensing and task-oriented sensing for edge AI, respectively. Finally, we emphasize interesting research directions on edge perception to motivate future works.

\end{abstract}

%

\vspace{-0.2cm}

\section{Introduction}\label{sec:intro}

Amidst the evolving landscape of information and wireless technology, {\it sixth-generation} (6G) network is envisioned to enable diverse intelligent applications, such as smart cities, smart healthcare, smart transportation, and digital twin \cite{CuiJSAC2023}, by providing high-accuracy sensing and high-throughput communication capabilities. For instance, robots installed with sensors are responsible for various tasks on the production line, such as assembling components, inspecting product quality, and packaging. 
In particular,  during the production process, a vast amount of data needs to be collected to enable monitoring alerts and rapid localization, providing effective decision support to streamline manufacturing processes.
Therefore, it is crucial to accurately sense environmental objects, swiftly analyze sensing data, and then make real-time decisions. 
Recently, {\it Integrated sensing and communication} (ISAC) has emerged as a new design paradigm that combines sensing and communication functions in 6G wireless networks \cite{FanLiu-JSAC}.  It thus gives rise to the wireless perceptive network that supports networked sensing \cite{ZhangVTM2021}, endowing multiple network  nodes  (such as {\it base stations} (BSs) and mobile terminals) with advanced sensing capabilities without relying on external sensors. 

However, future network extensively uses various sensing and generates a vast amount of data, but lacks a well-established connection between the sensed data and subsequent computing procedures, thus leaving room for task-oriented sensing design. 
Additionally,  sensing with larger data volumes requires the participating of multiples devices with higher computational capabilities, and is greater sensitivity to latency, compared to traditional communication. For instance,  an autonomous vehicle equipped with 12 cameras, 9 millimeter-wave radars, and 1 LiDAR could generate approximately 2.3 GB of data per  second, while the decision-making process is typically required to be completed within 20 milliseconds for safety.
Nevertheless, current communication networks often prefer to transmit data to remote (cloud-based) data centers for processing, which is not suitable for the existing sensing data streams. Therefore, the sensing and computing units of future network should primarily be deployed at the network edge. 
Edge-based computing solution enables prompt data access due to the proximity between the computation capacity and data resource at the network edge.
It thus promotes the emerge of edge {\it artificial intelligence} (AI), by pushing AI abilities to network edge, leading to a significantly reduced response delay with an efficient and privacy-aware inference assurance, aligning with the demands of real-time and context-sensitive applications. 
Consequently, it naturally gives rise to an interesting question: whether sensing and computing could be integrated at network edge, or even perform mutual benefits?




\begin{figure*}
\centering
 \setlength{\abovecaptionskip}{-1mm}
\setlength{\belowcaptionskip}{-1mm}
    \includegraphics[width=6.6in]{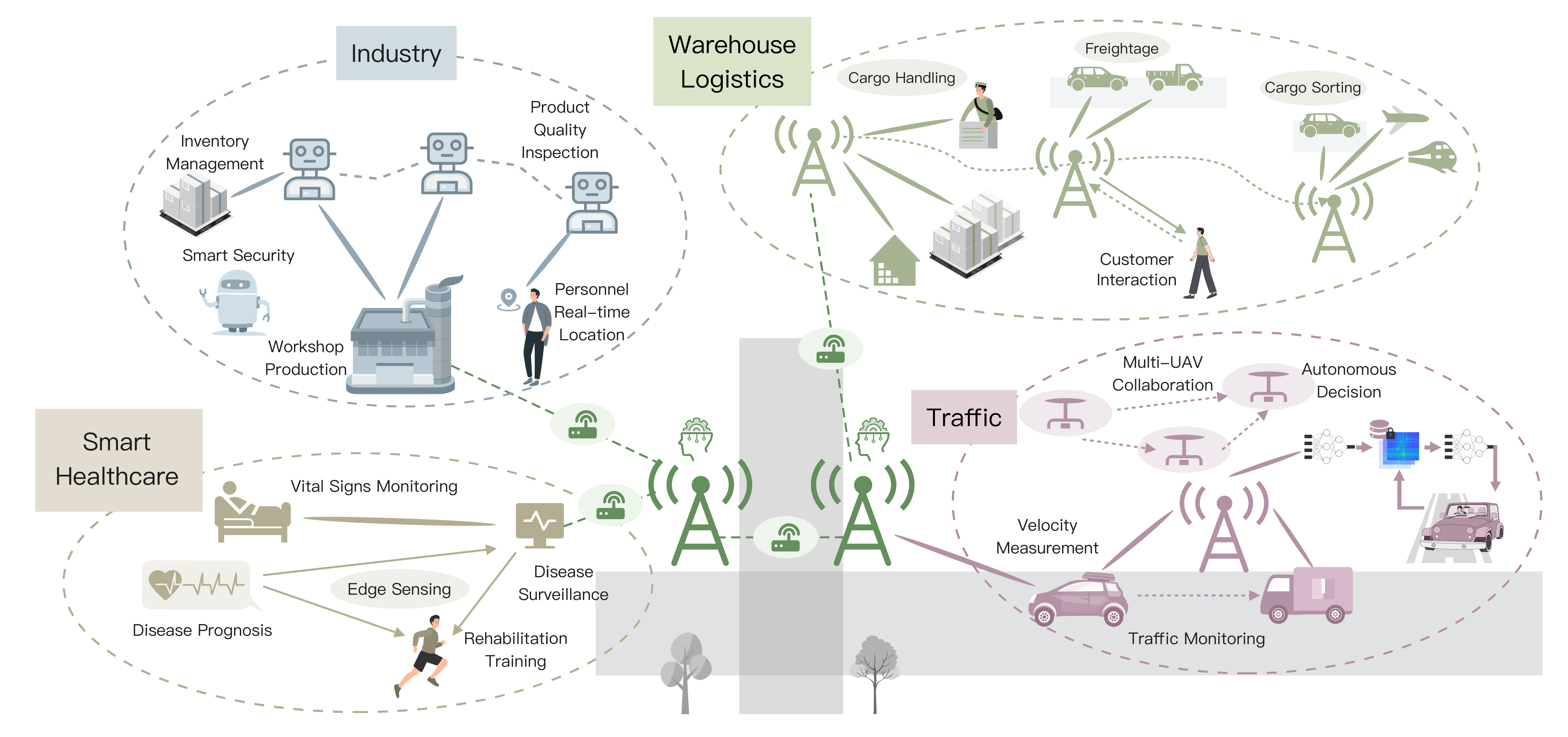}
\caption{Illustration of applications in edge perception. } \label{fig:Application}
\vspace{-0.2cm}
\end{figure*}

The computing resources of sensing functionalities are mainly distributed at network edge, instead in the cloud or on the end devices, which results in \textit{edge perception} to the next generation network, enabling real-time adjustments, rapid respond, and privacy protection.
In terms of wireless sensing, more design freedom on air-interface could be achieved by jointly scheduling cross-devices wireless resources in edge perception network. For instance, wireless data often suffers from quality issues occurred by interference and noise. By moving the processing of sensed data to the network edge, it could quickly re-design the air-interface based on computation results for high-quality data acquisition, or even achieve integrated end-to-end design through AI for data redundancy reduction.
It empowers the edge network to directly modify sensing functionality in real time, leading to optimized resource allocation, thus fostering new applications as shown in Fig. \ref{fig:Application}.

In view of the growing research interests, this paper provides a comprehensive overview of edge perception, which incorporates wireless sensing, communication, computation, and  AI capabilities at the network edge. In the following, we firstly delineate the data processing pipeline across physical, network, and application layers, which generally includes transmit and receive signal processing, network-wise cooperation, and application-specific data analysis. Then,  we present two paradigms regarding the interplay between edge AI and wireless sensing, i.e.,  edge AI empowered sensing and task-oriented sensing for edge AI, respectively, by particularly focusing on their challenges and potential solutions. Finally, interesting future directions on edge perception is emphasized. 

\section{Wireless Edge Perception}

\begin{figure*}
\centering
 \setlength{\abovecaptionskip}{-1mm}
\setlength{\belowcaptionskip}{-1mm}
    \includegraphics[width=6.6in]{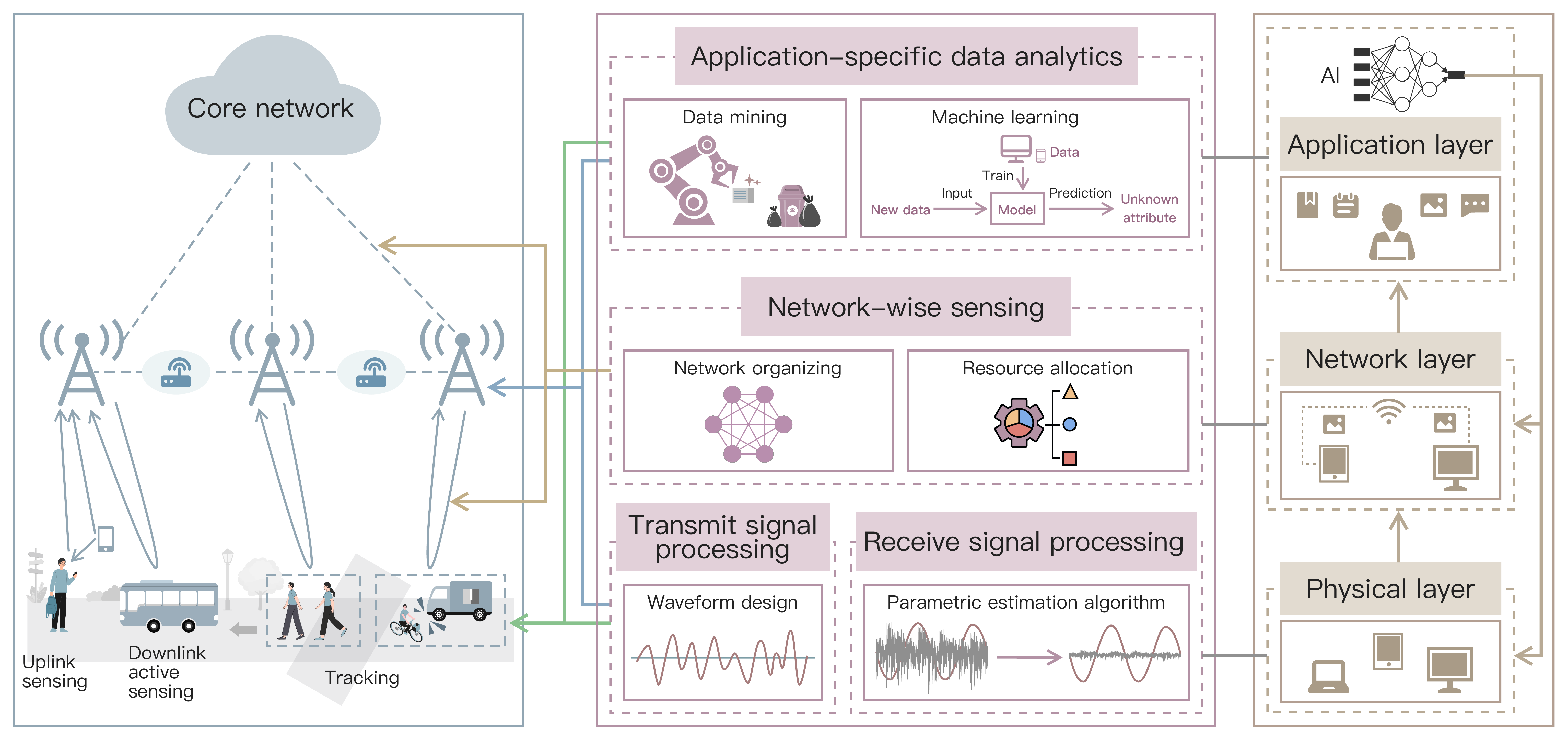}
\caption{Illustration of a pipeline of the edge perception procedure. } \label{fig:Sensing}
\vspace{-0.2cm}
\end{figure*}

In the following, the commonly concerned procedure of edge perception, as illustrated in Fig. \ref{fig:Sensing},  is discussed from the perspectives of physical, network, and application layers, by particularly focusing on transmit and receive signal processing, network-wise cooperation, and application-specific data analysis.

\subsection{Transmit Signal Processing}

Conventional wireless sensing focuses only on the receive signal processing design based on mostly {\it frequency modulated continuous wave} (FMCW) signal. 
 By pushing sensing and computing to the network edge, the increased freedom in jointly designing transmit signals from multiple sources represents a natural evolution.

\begin{itemize}
\item \textbf{Transmit Signal Design}: There are three transmit signal design strategies, namely sensing-centric, communication-centric, and joint communication-sensing designs \cite{FanLiu-JSAC}. 
Due to the demand of edge AI on the high-quality data samples, sensing-centric densign needs to allow dynamic adjustment to the transmit waveform relying on the pre-processing results of sensed data. Communication-centric design prioritizes communication performance and seeks to improve sensing quality by catering the sensed results. 
Considering various design metrics (such as range resolution and beampattern for sensing, as well as throughput and latency for communication), joint design provide a potential performance balance of current priorities between sensing and communication, where signals can be jointly designed under the pre-processing data results.

\item \textbf{Passive Sensing and Active Sensing}:  
Due to complicated environment and divers applications, sensing paradigm shifts from passive to active sensing. Passive sensing modules can collect data without emitting any signals, or detect naturally occurring signals, such as video cameras. Conversely, active sensing modules such as wireless sensing can adapt to surrounding environment by emitting radio frequency signals. Hence, properly designing the transmit waveform determines how the system senses the environment, which is reﬂected by the characteristics of the ambiguity function \cite{Guey-TIT}. This thus aids to easily extract additional information from surroundings. Therefore, designing an appropriate waveform can enhance the sample quality for neural networks, thereby accelerating the subsequent learning process.


\item \textbf{Cooperative Sensing}: Cooperative sensing would be a key scenario, especially among multiples nodes (e.g., BSs).  
In this case, many BSs/terminals are coordinated to observe the same target environment while supporting uninterrupted connectivities for communication devices.  Although these cross-node signals would cause interference thereby compromising the communication performance, it can be exploited as a useful factor to facilitate networked sensing \cite{Cheng2024TWC}. Specifically, the interference can be utilized for cooperatively performing networked sensing via increasing the sensing signal strength. Towards this end, the design of coordinated beamforming becomes necessary to provide joint sensing and communication capabilities.
 
 \end{itemize}

\subsection{Receive Signal Processing}

At the receiver side, it focuses on various aspects for efficient sensing, as illustrated below.
\begin{itemize}
\item \textbf{Joint Transmitter-Receiver Design}: 
Transmitter and receiver can always be jointly designed to maximize sensing performance at the network edge, in an efficient and convenient manner. This means that there is no need to send target information back to the transmitter.
Although it could help optimize the signal transmission process, thus facilitating efficient information extraction, the existing transceiver designs are mainly invoked to balance the sensing and communication capabilities. In this case, the consideration on AI processed results becomes a crucial factor, particularly in the scenario of using neural networks to complete the transceiver design for edge sensing. 
In specific, it can use neural networks to automatically optimize the signal transmission and reception processes without relying on pilot signals \cite{AoudiaTWC2022}.

\item \textbf{Monostatic and Bistatic Sensing}: Monostatic sensing approach utilizes the target echo signals to determine the distance, speed, and other characteristics of the target, while bistatic sensing exploits the target-reflected signals from a separate transmitter to extract the {\it channel state information} (CSI) and understand the target  characteristics in the environment for the purpose of estimation or classification. 
In most cases,  monostatic and bistatic methods can be jointly exploited to provide a more robust and flexible sensing solution for the receiver, as combining CSI data and target echo data can improve the accuracy of target detection and environmental sensing.



\item \textbf{Clutter and Interference Removal}: The undesired clutter from other devices in the surroundings and coupled echos from multiple targets introduced by receiver deployment, would severely degrade the quality of received signals. 

\item \textbf{Clock Asynchronism }:
The clock asynchronism between transmitters and sensing receivers that are spatially separate would affect the signal demodulation and sensing results. 

\item \textbf{High-Resolution Data}:  High-resolution types of sensing data (such as video), each with unique characteristics and processing requirements on data acquisition and features extraction, also present a challenge on the algorithm design.

\end{itemize}


\subsection{Network-wise Sensing}


Sensing at the network edge exploits the existing network structure to perform coordination between multiple potential sensing devices, e.g. sensing terminals, ISAC basestations, etc.\cite{ZhangVTM2021,Cheng2024TWC}.
 Meanwhile,  the 5G-A and 6G networks will achieve centimeter-level positioning accuracy, and deploy network-native AI technology, which naturally form a basis to the edge perception.  Nevertheless, sensing at the edge of cellular network still encounters several challenges. 
\begin{itemize}
\item  \textbf{Prior Knowledges}: The environment for networked sensing is inherently more intricate due to potential interference between communication and sensing activities. As such, acquiring proper  prior knowledges of the site-specified electromagnetic propagation environment, becomes crucial for achieving high-precision perception.  For instance, though the deployment of dense BSs would not be easily changed in site-specified cellular networks, the pre-stored prior maps could greatly improve the beam management and signal processing efficiency. Therefore, how to make good use of these prior knowledges is of significance for facilitating edge sensing tasks.
\item  \textbf{Real-time Sensing}: The requirement of real-time sensing process introduces additional complexities into edge perception networks, resulting in increasing data transmission and processing delays, especially in the presence of moving distributed targets \cite{Nie2024aa}. One possible solution is to design a fronthaul-based approach to coordinate multiple sensing nodes within a specified space, with data being collected by the {\it antenna array unit} (AAU) and then aggregated in the {\it baseband unit} (BBU).  Additionally, another design freedom in edge perception networks is the spatial correlation among neighboring BSs/devices in space, which efficiently affects the way data is collected, transmitted, and processed. In specific, the deployment of devices can be optimized according to the correlation for reducing redundant data collection and transmission.
\end{itemize}




\subsection{Application-specific Data Analytics}


Unlike the communication data that places emphasis on the transmission efficiency and reliability, sensing data are required to be processed and analyzed in perception networks for further decision making and adjustment, especially facing two unique types, namely short-package and multi-modal data.

\begin{itemize}
\item \textbf{Data Characterization}: After pre-processing, sensed data transmitted over cellular network may usually be burst, short-package, or extensive, such as vehicle location report in smart transportation system. Meanwhile, multi-modal data (such as images, videos, sensor readings, and text) is generated from different sources to gain a more comprehensive understanding of environment, like the mmWave radar and camera in autonomous driving.  Both of them meet the requirements of demands on the rapid response and data simplification and diversity, on the purpose of facilitating follow-up decision making. In this case, how to ensure or distill the insightful value of these data is a challenge for data analysis. 

\item \textbf{Processing Position}: Different positions (e.g. AAU/on-device or BBU or  edge server) lead to different levels of data processing and analysis, such as raw data collected by devices, pre-processed at BBU, analyzed at server. For example, raw data processing on edge devices could handle some naive task-oriented application-specific tasks, such as feature extraction and pattern recognition. 
Worthy to note that data processing at different positions would cause uneven delay, depending on transmission distance and the performance of different units, such as {\it central processing units} (CPUs) and {\it graph processing units} (GPUs), in terms of their processing speed and efficiency. Besides,  processed sensing data typically incurs some information loss, particularly through data compression and filtering techniques.
Thus, it is crucial to choose the sensing data processing positions for balancing processing delay and information loss. 

\item \textbf{Task-specific Data Processing}:
Data processing is inherently task-specific and application-oriented. Different applications have different requirements on the physical layer. Taking autonomous driving as an example, its sensing data consists of both multi-modal and short-package types with a relatively large scale. Therefore, higher requirements are placed on the design of transceivers, requiring not only high bandwidth and strong data collection capability but also enhanced processing capability for individual vehicles (such as carrying AI chips).
\end{itemize}





\section{Edge AI Empowered Sensing}

\begin{figure*}
\centering
 \setlength{\abovecaptionskip}{-1mm}
\setlength{\belowcaptionskip}{-1mm}
    \includegraphics[width=6in]{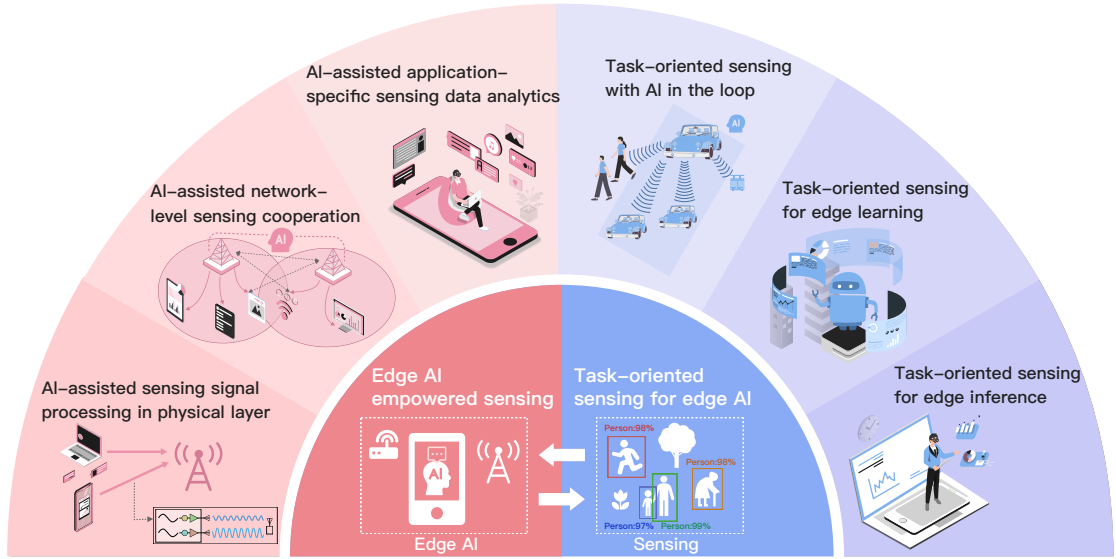}
\caption{Illustration of the interplay between edge AI and wireless sensing. } \label{fig:AIS}
\vspace{-0.2cm}
\end{figure*}

Following the above bird's-eye view on sensing at network edge, edge AI will clearly perform a vital role in enhancing wireless sensing capabilities across physical, network, and application layers, as depicted in Fig. \ref{fig:AIS}.
In this section, we showcase how edge AI contributes to wireless sensing across physical, network, and application layers.

\subsection{AI-assisted Sensing Signal Processing in Physical Layer}

In general, sensing tasks often occur in dynamic and unpredictable environments. By incorporating AI algorithms at the edge, devices can autonomously learn and adapt system designs based on real-time environmental information.

\begin{itemize}
\item \textbf{AI-assisted Transceiver Design}: 
Emerging AI technologies such as {\it deep learning} (DL) demonstrate strong data-driven capabilities in various wireless communication applications, ranging from channel estimation, signal detection, resource allocation, and more. They are usually adopted to dig out and harness valuable data features to enhance task performance even further. For instance, in a vehicle-road coordination scenario,  multi-modal surroundings data (recorded by GPS, LiDAR, camera, and more) are collected to be processed at edge server to aid the beamforming design by leveraging DL methods. Besides, implementing learning-based algorithms could help efficiently distill useful sensing insights from received signals, which are subject to rich clutter from the complex propagation environment in edge perception networks. For example,  a learning based parameter estimation scheme could enhance the accuracy of delay and  angle-of-arrival estimations and provide robustness to the channel reconstruction error \cite{LushanurTAES2020}.

\item \textbf{Resource-Friendly Approach}: 
Note that edge devices often have constrained computation and energy resources, making it inefficient to implement AI models for real-time physical layer optimization. One way to tackle this issue is to deploy lightweight AI models at edge devices, which are optimized with fewer trainable parameters and layers. Taking MobileNet as an example, the preprocessed signal data is fed into the trained model at edge devices to identify interference sources. Moreover, {\it neural processing units} (NPUs) or AI chips, are increasingly integrated into edge devices, such as Apple Neural Engine. These specialized chips are designed to efficiently handle the computation-intensive tasks, making them ideal for real-time applications in constrained environments.  
\end{itemize}

\subsection{AI-assisted Network-level Sensing Cooperation}

Aiding with intelligence, edge perception network could leverage more sophisticated algorithms for more accurate sensing. Particularly, network-level sensing scenario enables extensive sensing collaboration across multiple nodes.  Actually, the evolution of network, from single-cell to network-level and potentially cell-free architectures, is observed to obey the characteristics of intelligence and collaboration. In single-cell network, sensing performance is constrained by the capacity and coverage of individual node, while collaboration among sensing nodes (BSs or devices) significantly enhances perception range and precision in network-level sensing. Ultimately, the cell-free architecture will break down traditional cell boundaries, allowing all nodes to collaborate freely under intelligent scheduling, providing seamless and efficient sensing services. 



Current networks are evolving towards more integrated and intelligent systems, which are characterized by several key features, including heterogeneity,  collaboration between different entities, and distributed data collection, as elaborated in the following.
\begin{itemize}
\item \textbf{Heterogeneity}: Note that next generation network is usually heterogeneous, within which macro and micro BSs (such as femtocell, picocell, and even WiFi access points) are both involved and their capabilities on sensing, communication, and computation are uneven.  In this case, their collaboration indicates a way to achieve a dual-function balance of sensing and communication at the network level. Utilizing learning-based algorithms could efficiently organize the network-level collaboration with respect to any specific sensing tasks to achieve well-balanced performance tradeoff.

\item \textbf{Sensing Cluster among Different Entities}: To better manage the sensing traffic, how to select sensing nodes (BSs, devices, or sensors) and construct cooperative clusters is necessary.  
As current network is designed for communication, sensing functionality cannot work in some scenarios. Therefore, scheduling multiple sensing nodes to form  sensing clusters can enhance the sensing accuracy in a specific area, where raw sensing data can be shared, complemented, and combined \cite{JiComMag2023}. 
This collaborative approach thus enhances the sensing effectiveness, enabling more comprehensive and accurate data processing. In this case, neural networks could be employed to evaluate the conditions of each node candidate and thus adapt scheduling policies accordingly.

\item \textbf{Distributed Data Collection}: The collaboration of distributed network nodes provides a broader coverage area to provide more extensive environmental data collection with higher accuracy and granularity. However, it also introduces challenges in maintaining continuous communication and synchronization across dispersed nodes. Deploying AI models would offer a robust approach in managing handover and synchronization without any sophisticated prior information of channel. 
\end{itemize}




\subsection{AI-assisted Application-specific Sensing Data Analytics}

In edge perception networks, the collected wireless sensing data (i.e., CSI or received waveforms) would be preprocessed to remove clutter and noise, from which we could extract task-related features to achieve any specific perceptual task.

There exists a common concern  of embedding DL models into wireless sensing systems, i.e., the DL techniques rely on massive high-quality data.  The increasing complexity of tasks necessitates larger data volumes, higher quality, and learning parameters to be configured. Due to the frequent occurrence of loss, redundancy, mislabelling, and class imbalance in sensed data, the scalability and generalization of training and learning processes are significantly hindered. Taking Wi-Fi sensing as an example,  its collected data cannot be labelled automatically, which are heavily affected by the surrounding environments with multi-path signal propagation.  

One possible solution is to first use advanced signal processing for extracting definite components of wireless signals and then reconstruct DL algorithms for specific sensing tasks, namely teaching the training network to learn from multi-dimensional information with well-designed loss functions \cite{LiACM2022}. Besides,  data augmentation technique is another promising approach to tackle this challenge, by generating  extensive virtual data samples through {\it generative adversarial networks} (GANs),  which is made up of two components, namely a generator for generating samples, and a discriminator attempting to distinguish these samples from real data.

\section{Task-oriented Sensing for Edge AI}

The integration of wireless sensing and ISAC also brings substantial advantages for edge AI to support in-network data acquisition, swift processing, and rapid decision making. For example, sensors installed on vehicles can consistently capture a wide array of road-related information encompassing vehicle movements, pedestrian activities, and road conditions. These data can be swiftly collected by the edge network, facilitating immediate on-device training and prediction in autonomous driving systems.  Furthermore, at network edge, sensing devices are typically dispersed across varied environments, affording them the capability to capture a diverse range of data types. Such a diversity aids AI models in better understanding and adapting to varying circumstances and conditions, thereby improving the robustness and adaptability of AI systems.  Furthermore, the edge network parameters could be configured by utilizing sensed data to adapt the characteristics of varied environmental conditions or network demands.  In this section, we concentrate on the recent advancement on edge AI with sensing in the loop, from three perspectives of task-oriented sensing, edge learning, and edge inference, respectively.

\subsection{Task-oriented Sensing with AI in the Loop}

Multiple sensors/devices prefer to monitoring the physical world  based on the specific requirements of a given perception task, thus enabling the so-called \textbf{task-oriented sensing in edge perception network}.  Unlike traditional sensing that samples and transmits all packets in a fixed way, task-oriented sensing strategically samples crucial and relevant information with higher priority by adaptively adjusting the sensing area or sampling rate, thereby effectively reducing data redundancy.  In this approach, the system allocates resources such as sensing time and sensing energy more efficiently, focusing on {\it field of view} (FoV) and thus optimizing the overall sensing process. The crux of achieving this lies in characterizing AI task properties and guiding the design of sampling and scheduling processes. However, how to determine the property and their quality of service based on the specific task is an open problem.  Existing methods to measure this property are usually based on the freshness, error, and context of information, such as age of information, mean squared error, and urgency of information.
Additionally, a sensing data platform for intelligence, sensing, and communications has been established to facilitate the training of large DL models for sensing (see \url{www.sdp8.org}).

\subsection{Task-oriented Sensing for Edge Learning}

   \begin{figure}
\centering
 \setlength{\abovecaptionskip}{-1mm}
\setlength{\belowcaptionskip}{-1mm}
    \includegraphics[width=3.7in]{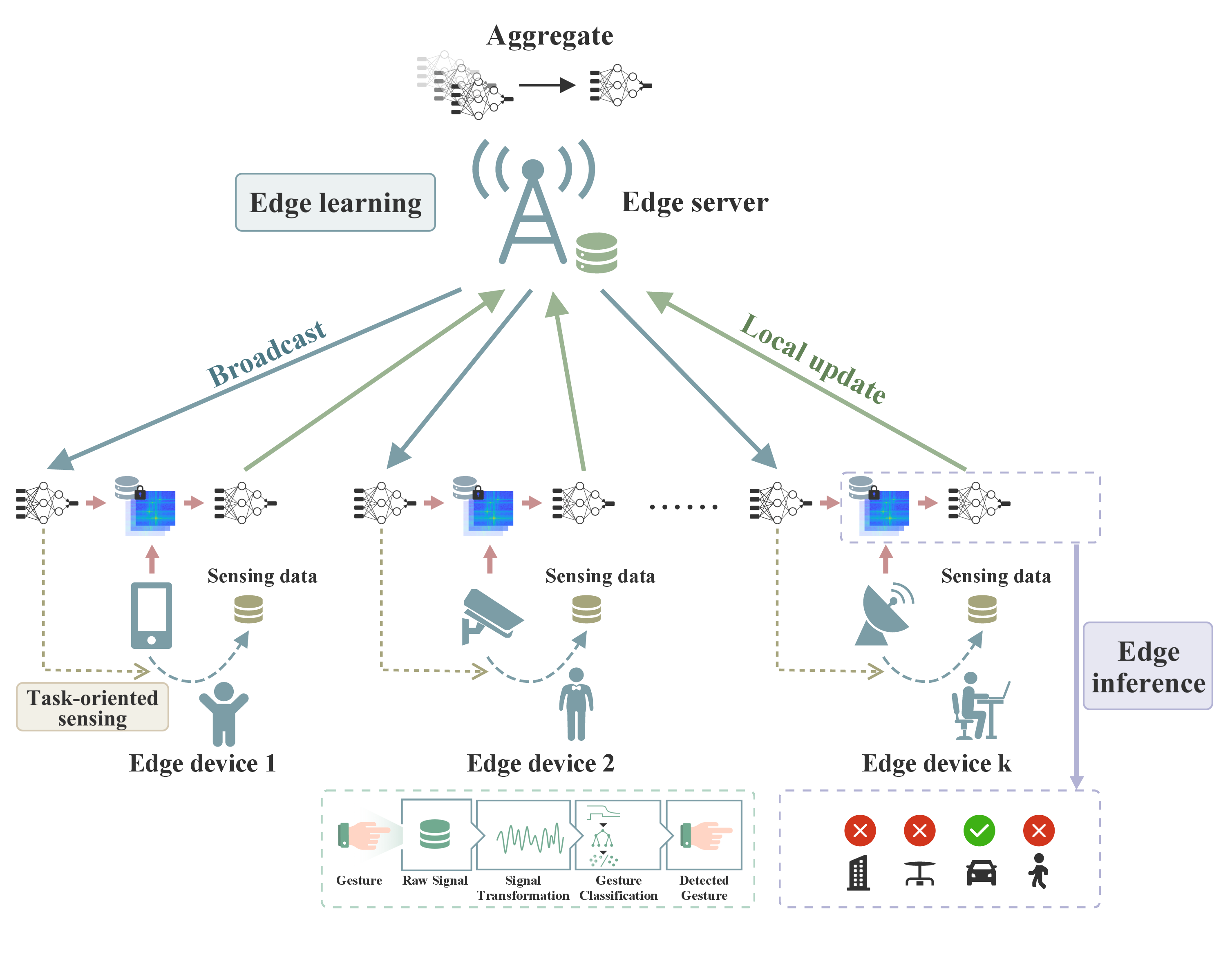}
\caption{Illustration of edge AI network. } \label{fig:EdgeLearning}
\vspace{-0.2cm}
\end{figure}

In edge learning, edge devices are coordinated to train a shared AI model, especially the sensing samples would be collected via wireless sensing.  Take {\it federated edge learning} (FEEL) as a classical example.

As shown in Fig. \ref{fig:EdgeLearning},  FEEL is implemented using distributed stochastic gradient descent  to minimize a well-designed loss function by optimizing AI-model parameters in an iterative way while preserving sensed data locally. 
In specific, integrating sensing into FEEL workflow introduces a consideration on the need for newly sensing data. Such continuous input of data ensures that the learning model is trained on the most recent and relevant information, reflecting real-time environmental changes. Also, taking the data importance into account, like   sensitive area or objects, the learning process becomes efficient as the redundancy of training data could be reduced via controlling the sampling frequency.

Note that existing edge networks are designed as separated processes of sensing, communication, and computation, each tailored for distinct objectives.  
Therefore,  it prompts a more integrated and holistic approach of  {\it integrated sensing, computation, and communication} (ISCC) to fully unleash the potential of edge perception network for better providing in-network data processing and seamless AI services.
To demonstrate the advantage of the proposed ISCC network, a concrete example of human motion recognition is provided in \cite{Zhu2023SCIS}. In particularly,  multiple edge devices collect data for human motion recognition via wireless sensing and communicate with an edge server to exchange model updates in an ISCC system.  Each device updates its local model using the most recent sensing data samples, with the batch size determined by the quantity of samples. 
However, existing work on ISCC only considers the loose connection among sensing, communication, and computation, which means that the sensing data only directly affect the learning samples instead of learning performance. Hence, building a tight relationship between sensing and learning based on ISCC architecture to reveal how the sensed data influence learning procedure and then inversely adjust sensing process is still unknown.



\subsection{Task-oriented Sensing for Edge Inference}


In addition to the edge learning, edge inference is another crucial aspect that plays a pivotal role in the successful implementation of edge AI.  In specific, edge inference refers to the process of deploying and executing a well-trained AI model directly at the network edge to achieve different goals.

Depending on the location of AI-model deployment, there are three different methods to implement edge inference, namely on-device inference, on-edge inference, and split inference or edge-device co-inference. For on-device inference, a lightweight AI-model is deployed at edge devices, where sensed data at single node is locally processed for inference execution in real time. 
However, due to the substantial computational workload, applications based on {\it deep neural networks} (DNNs) cannot be fully executed due to their low-performance computing units and limited battery life. 
Thus, {\it on-edge inference} pushes the sensed data being delivering from edge devices to a separate edge server (e.g., a BS), where DNN models are fully deployed.  Enjoying the data fusion from distributed sources, it provides a more comprehensive and accurate understanding of the environment during the inference phase.  Nevertheless, the substantial volume of sensed data (such as high-definition videos, and point clouds) from terminals may pose significant communication challenges, leading to a considerable communication overhead.


To tackle the above issues, edge-device co-inference under the collaboration between edge server and devices  is proposed based on split inference by separating the AI model into two submodels. One is deployed on edge devices, primarily responsible for feature extraction tasks, while the other operates at the edge server to handle the remaining inference tasks. 
Consequently, it offers privacy preservation by designing task-specific sensing and eliminating the need for transmitting raw data, and lowers hardware demands by shifting intensive computational tasks to edge server \cite{LiTWC2020}. 
Nonetheless, because of the inability to efficiently correlations between features extracted from various devices, the communication overhead and latency emerges as a major bottleneck for multi-device collaborative edge inference. 
Hence, a new multi-device edge inference system that leverages the multi-modal data collected by collaborative sensing among devices to extract more insightful  features has investigated  in  \cite{Wen-JSAC} to implement low-latency intelligent services. 


%
%
%

\vspace{-0.2cm}

\section{Conclusion}

This article presented a comprehensive overview on edge perception that incorporates the sensing, communication, computation, and AI functionalities at the network edge.  We discussed wireless edge perception, and delineated the evolutionary progression of edge perception network from the interplay between edge AI and wireless sensing. It is our hope that this article can provide new insights on this interesting research topic, and motivate more interdisciplinary research from the communities of wireless communications, machine learning, and wireless sensing.

This work also opens several directions for further investigation, as briefly discussed in the following.
\begin{itemize}
\item Generative AI as an emerging and promising technology plays a significant role in addressing challenges related to data processing networks \cite{Du2023aa}. First, for multi-modal sensing data in edge perception networks, generative AI could generate synthetic data based on the available modalities, and augment existing datasets with limited labeled data, thus improving robustness and generalization. While for those fragmented sensed data due to the short package property, it can create realistic samples under observations, and accordingly predict the sequences, making a decision on the next data points.
\item {\it Large language models} (LLMs) are advanced AI models working well in understanding and generating human-like text.  These models, such as {\it generative pre-trained transformer} GPT series, are capable of performing a wide range of natural language processing tasks, including language translation, text summarization, question answering, and text generation. Leveraging LLM  is expected to offer more comprehensive knowledge support for physical information systems in edge perception network \cite{Xu2023aa}. 

\end{itemize}

\bibliography{AirCompforFL}
\bibliographystyle{IEEEtran}

\vspace{-2.2cm}

\end{document}